\documentclass{article}

\usepackage{arxiv}

\usepackage[utf8]{inputenc} 
\usepackage[T1]{fontenc}    
\usepackage{lmodern}        
\usepackage{hyperref}       
\usepackage{url}            
\usepackage{booktabs}       
\usepackage{amsfonts}       
\usepackage{nicefrac}       
\usepackage{microtype}      
\usepackage{lipsum}
\usepackage{graphicx}

\title{Using automated decision-making (ADM) to allocate Covid-19
vaccinations? Exploring the roles of trust and social group preference
on the legitimacy of ADM vs.~human decision-making.}

\author{
    Marco Lünich
    \href{https://orcid.org/0000-0002-0553-7291}{\includegraphics[scale=0.06]{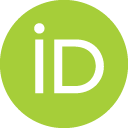}}
   \\
    Department of Social Sciences \\
    Heinrich Heine University \\
  Düsseldorf, Germany \\
  \texttt{\href{mailto:marco.luenich@hhu.de}{\nolinkurl{marco.luenich@hhu.de}}} \\
   \And
    Kimon Kieslich
    \href{https://orcid.org/0000-0002-6305-2997}{\includegraphics[scale=0.06]{orcid.png}}
   \\
    Department of Social Sciences \\
    Heinrich Heine University \\
  Düsseldorf, Germany \\
  \texttt{\href{mailto:kimon.kieslich@hhu.de}{\nolinkurl{kimon.kieslich@hhu.de}}} \\
  }

\newlength{\csllabelwidth}
\newlength{\cslhangindent}
  {}%
  {\par}
\newenvironment{CSLReferences}[3] 
 {
  \setlength{\parindent}{0pt}
  \ifodd #1 \everypar{\setlength{\hangindent}{\cslhangindent}}\ignorespaces\fi
  \ifnum #2 > 0
  \setlength{\parskip}{#2\baselineskip}
  \fi
 }%
 {}
\usepackage{calc} 

\usepackage{booktabs}
\usepackage{longtable}
\usepackage{array}
\usepackage{multirow}
\usepackage{wrapfig}
\usepackage{float}
\usepackage{colortbl}
\usepackage{pdflscape}
\usepackage{tabu}
\usepackage{threeparttable}
\usepackage{threeparttablex}
\usepackage[normalem]{ulem}
\usepackage{makecell}

\begin{document}
\maketitle

\def\tightlist{}

\begin{abstract}
In combating the ongoing global health threat of the Covid-19 pandemic,
decision-makers have to take actions based on a multitude of relevant
health data with severe potential consequences for the affected
patients. Because of their presumed advantages in handling and analyzing
vast amounts of data, computer systems of automated decision-making
(ADM) are implemented and substitute humans in decision-making
processes. In this study, we focus on a specific application of ADM in
contrast to human decision-making (HDM), namely the allocation of
Covid-19 vaccines to the public. In particular, we elaborate on the role
of trust and social group preference on the legitimacy of vaccine
allocation. We conducted a survey with a 2x2 randomized factorial design
among n=1602 German respondents, in which we utilized distinct
decision-making agents (HDM vs.~ADM) and prioritization of a specific
social group (teachers vs.~prisoners) as design factors. Our findings
show that general trust in ADM systems and preference for vaccination of
a specific social group influence the legitimacy of vaccine allocation.
However, contrary to our expectations, trust in the agent making the
decision did not moderate the link between social group preference and
legitimacy. Moreover, the effect was also not moderated by the type of
decision-maker (human vs.~algorithm). We conclude that trustworthy ADM
systems must not necessarily lead to the legitimacy of ADM systems.
\end{abstract}

\keywords{
    automated decision-making
   \and
    trust
   \and
    social group preference
   \and
    Covid-19
   \and
    legitimacy
   \and
    public opinion
  }

\hypertarget{introduction}{%
\section{Introduction}\label{introduction}}

The spreading novel coronavirus SARS-CoV-2 and the ongoing global
COVID-19 pandemic coincide with the worldwide proliferation of computer
technology in everyday life. Consequently, computer systems have also
been widely regarded as a viable instrument in combating the pandemic
(Bragazzi et al. 2020; Calandra and Favareto 2020; Jacob and Lawarée
2020; Malik et al. 2020; Nguyen et al. 2020; Sipior 2020). For instance,
aiming to mitigate the loss of life and to find treatment, cures, and
vaccines against the virus, the necessary medical research is
unthinkable without computers. Beyond their general use as research
instruments for medicine and public health, computer systems nowadays
are also seen as a potent and helpful tool in tackling the more social
issues of a pandemic. As a prime example, systems of automated
decision-making (ADM) have been deployed to automatically and fairly
prioritize persons for vaccination better to coordinate the vaccination
of the population against the coronavirus.

Because vaccination prioritization is a hotly debated social issue, and
the unreflected use of technology may come with severe social
consequences, this implementation of ADM receives particular public
scrutiny. In many cases where ADM systems were deployed, it became
quickly apparent that their decisions for prioritization were biased,
leading to backlash and outright rejection (Ciesielski, Zierer, and
Wetter 2021; Guo and Hao 2020).

Nevertheless, even if ADM systems consistently followed a formally fair
algorithm as intended by their makers, the public may still question the
algorithmic systems' decisions despite their formal attainment of
optimization goals. After all, algorithms may arrive at optimized
decisions that correspond to formally correct and fair outcomes but are
unintuitive to a lay public as decisions may entirely oppose people's
social preferences and moral ideas. Negative evaluations of
controversial public decision-making pose a general social problem --
regardless of whether those decisions are based on human decision-making
(HDM) or ADM. Notwithstanding, the question arises whether the type of
decision-maker (ADM vs.~HDM) influences the decision evaluation.

To shed light on this issue, in this paper, we first ask to what extent
ADM is perceived as a viable solution for the distribution of the
vaccine and examine the role of trust as an explaining factor for
viability perceptions. Second, we investigate the impact of decisions
concerning the prioritization in the coronavirus vaccine allocation on
perceptions of the legitimacy of the decision. In particular, we examine
the consequences for decision legitimacy when decisions are unpreferred.
Contrasting such perceptions concerning ADM to a situation where humans
decide about prioritization, we also inquire whether the trust in the
agent making the decision moderates the supposed relationship between
the favorability of a decision and its legitimacy perception and whether
the proposed mechanisms differ between the two decision-making agents.

Drawing a quota sample from a German online access panel, results
indicate ambivalence in the general perception of ADM as a viable tool
for disseminating vaccines against the coronavirus among the German
population. However, higher general trust in ADM systems is positively
related to a more favorable assessment of the viability of their use in
vaccine distribution. Using a factorial survey design that randomly
varies the prioritization of different social groups (prisoners
vs.~teachers) in vaccine distribution and the agent deciding such
prioritization (ADM vs.~HDM), results also suggest that decisions that
assign a higher priority to unpreferred groups are perceived to be less
legitimate. Contrary to the authors' expectations, the trust in the
agent making the decision did not moderate this relationship, and there
is also no difference between ADM and HDM concerning a moderating effect
of trust.

As ADM systems may have adverse and especially discriminating
consequences and the use of ADM systems hinges on widespread public
acceptance, the resulting insights into the determinants of public
support concerning ADM provide valuable information regarding their
implementation. We consequently discuss implications for executives,
politicians as well as actors from civil society.

\hypertarget{adm-systems-in-prioritization-of-the-coronavirus-vaccine-distribution}{%
\subsection{ADM Systems in Prioritization of the Coronavirus Vaccine
Distribution}\label{adm-systems-in-prioritization-of-the-coronavirus-vaccine-distribution}}

The societal distribution of limited goods, such as medical resources
and especially vaccines, is a social challenge that warrants research
attention (Grover, McClelland, and Furnham 2020; Huseynov, Palma, and
Nayga 2020; Ratcliffe 2000). The prioritization of vaccination is a
hotly debated public issue as the world faces the threat of a global
pandemic in early 2021. The roll-out of the international vaccination
program against the coronavirus saw itself confronted with a limited
amount of vaccine that needed to be distributed to the human population
as rapidly and effectively as possible.

As a result, such a vaccine distribution process often relies on many,
especially multi-faceted data points from patients, namely their age,
work occupation, or pre-existing health issues determining an individual
risk status that leads to the decision about (non-)prioritization of a
person. The rule-based distribution then usually technical formulations
that structure and evaluate such input data according to pre-determined
distribution criteria for assigning vaccines.

The more data points considered and the more sophisticated the
allocation formula, the more difficult it becomes for human
decision-makers to assess whom to vaccinate first and to establish an
order for vaccination. Consequently, computer systems have been deployed
to assist in managing this process. Moreover, formalized algorithms and
computer systems can be used in the pre-determined vaccination
distribution organization and thus providing guidance in identifying and
implementing better-optimized distributions. In a simulation study,
``using an age-stratified mathematical model paired with optimization
algorithms,'' (Matrajt et al. 2020, 1) a research group shows how
different optimizing strategies lead to different recommendations on
whom to vaccinate first.

In theory, if one aims for a fine-grained allocation based on extensive
data-processing, digital tools may better optimize the allocation of
vaccines and do so more quickly. Thus, an algorithm may also relieve
medical or administrative staff in times of crisis. Consequently, ADM
may be seen as a viable solution for allocating coronavirus vaccines --
at least when it comes to the bureaucratic perspective of public
management and administrative decision-makers (Wirtz and Müller 2018).

In practice, despite great hopes for better outcomes, ADM systems have
often not been able to protect what appear to be the most vulnerable
groups and have led to unintended and morally questionable decisions.
Deployed as a tool to prioritize people for vaccination against the
coronavirus, ADM systems, too, have shown to produce incorrect and
biased decisions that have been regarded as morally wrong and unfair.

When an algorithm was tasked to distribute the first batch of vaccines
against the coronavirus at the Stanford Medical Center in the US in
December 2020, only a few physicians from the front lines were
prioritized (Guo and Hao 2020). While not all reasons for this results
are entirely public, a report by the \emph{MIT Technology Review}
highlights that the inclusion of age of employees was critical, since it
prioritized old and young staff. However, according to the report,
``frontline workers {[}\ldots{]} are typically in the middle of the age
range'' (Guo and Hao 2020). Secondly, another criticism was, that to
``expose to patients with covid-19'' (Guo and Hao 2020) was not included
as a factor. The resulting algorithm's preference for administrators or
doctors working from home resulted in a backlash against the ADM system,
protests from the hospital's residents, and quite some public attention
(Wu and Isaac 2020).

In the state of Bavaria in Germany in early 2021, an algorithm was used
to assign vaccination appointments to a pre-defined risk group that
consisted of people of 80 years or older as well as younger persons that
were assigned to the respective risk group due to having a high-risk
profile, e.g.~medical staff (Ciesielski, Zierer, and Wetter 2021).
Appointments prioritized people with a higher score based on their age.
However, the algorithm assigned a randomly chosen value between 80 and
100 to persons below 80 years of age. Consequently, the algorithm
discriminated against the younger octogenarians that were simply
assigned their true age and thus had a lower chance to receive an
appointment than younger people. The chance of being assigned randomly
to be older than 80 years is 95\% for the younger persons in the risk
group. As a result, only a few 80-year-olds received an appointment for
vaccination, causing complaints and extra effort and expenses as the
underrepresented group had to be manually contacted.

\hypertarget{adm-for-the-social-good-and-its-public-perception}{%
\subsection{ADM for the Social Good and its Public
Perception}\label{adm-for-the-social-good-and-its-public-perception}}

Such anecdotal evidence is in line with recent research that suggests
that the implementation of ADM in public administration is far from
unproblematic (Hartmann and Wenzelburger 2021). Even the most
well-intentioned ADM may falsely discriminate against certain groups,
i.e., such systems often violate the ``established weighting of relevant
ethical concerns in a given context'' (Heinrichs 2021, 1). These general
concerns regarding discrimination and biases have recently instigated
substantial research activity that addresses the social implications of
ADM implementation (Crawford et al. 2016).

To better guide the intricate development process of automated computer
systems for the `social good,' Berendt (2019) proposes four questions
that need to be asked in advance regarding the means and end of ADM
implementation: ``What is the problem {[}\ldots{]}?, Who defines the
problem?, What is the role of knowledge?, and What are important side
effects and dynamics?'' (p.~44). Accordingly, fighting the pandemic
threat of the coronavirus by distributing vaccines is a goal that
certainly benefits the social good and, as depicted above, may be
tackled using ADM. However, any attempt at appropriately answering
Berendt's questions reveals that ADM implementation may prove a complex
and intricate task in this regard. Depending on different assumptions
and preferences, approaches and results of ADM may vary extensively. For
example, depending on preference, the respective solution to the problem
of too many infections, or deaths, or too much economic damage, could be
defined as either ``lower case numbers (of certain groups),'' or ``lower
the death rate,'' or even ``ensure a fast return to normal life,'' or
all of the above. The problem can be defined by various stakeholders,
e.g., expert commissions, politicians, the media, or the broad public.
Then, as part of the knowledge problem, one must consider how problems
and solutions using ADM are framed by stakeholders via mediated public
communication and received and understood by all parties involved,
especially the public. Eventually, it is hard to determine important
side effects and unintended dynamics well in advance when it comes to
ADM.

Consequently, to better guide the implementation process and prevent
respective problems regarding the social good, the European Union offers
specific guidelines that include the demand to promote trustworthy ADM
as a solution for opaque and inaccessible applications. Thus, it becomes
clear that ADM's public perceptions are of utmost importance in the
intricate implementation process of ADM systems. For that reason, the
usage of ADM in combating the coronavirus pandemic serves as a prime
example. With a particular focus on the public perception regarding
questions of ADM discrimination and trust in the decision-making agent,
this raises important research questions that we investigate in this
study.

As a result, our paper adds to the pre-existing literature in three
important ways. First, it provides novel insights into the public
perception of ADM in combating the coronavirus. Second, it sheds light
on potential issues of ADM implementation, primarily when resulting
decisions are perceived as unpreferable. Third, it addresses the effect
of trust in agents making important decisions, contrasting human and
automated decision-making.

\hypertarget{the-perceived-viability-of-adm-in-the-distribution-of-vaccines}{%
\subsection{The Perceived Viability of ADM in the Distribution of
Vaccines}\label{the-perceived-viability-of-adm-in-the-distribution-of-vaccines}}

At the outset of addressing consequences by implementing ADM systems
stand the general public perception and assessment of the viability of
ADM in the distribution of vaccines. In general, expectations concerning
the use of ADM systems in decision-making include the decisions to be
quicker, more consistent, and in general more robust compared to human
decisions that are also expected to adhere to specific distribution
formulas (Dawes, Faust, and Meehl 1989; Kaufmann and Wittmann 2016;
Kuncel et al. 2013). Accordingly, in recent years, the use of ADM
systems in public administration has gained considerable traction (Wirtz
and Müller 2018) and -- as demonstrated by the two case examples from
the US and Germany above -- is also implemented for the distribution of
COVID-19 vaccines.

In general, people may be aware of the possibilities of computer systems
and their implementation regarding distribution processes, even if they
have not yet heard of specific ADM use cases, whether regarding vaccine
distribution or else. For instance, there is a general awareness
concerning the widely discussed impact of Artificial Intelligence (AI)
on society (Kelley et al. 2019). Strictly speaking, ADM may not
necessarily be AI. In terms of public perception, it still can be argued
that computer systems that autonomously make decisions are at least
associated with AI from a lay perspective (Cave, Coughlan, and Dihal
2019; Liang and Lee 2017). Research shows that many countries generally
have rather favorable attitudes towards AI (Kelley et al. 2019; Zhang
and Dafoe 2019). However, in specific contexts, consequential decisions
of AI may be perceived as threatening (Kieslich, Lünich, and
Marcinkowski 2021).

The first question of interest for our research is whether people
perceive ADM systems as a viable solution for distributing vaccines
against the coronavirus. We thus ask the following research question:

\emph{RQ1. To what extent do people consider ADM a viable solution for
the distribution of COVID-19 vaccines?}

\hypertarget{trust-in-algorithms-and-the-perceived-viability-of-adm}{%
\subsection{Trust in Algorithms and the Perceived Viability of
ADM}\label{trust-in-algorithms-and-the-perceived-viability-of-adm}}

ADM systems are often considered `black boxes' because it is often
impossible to make the inner workings of such systems transparent and
comprehensible (Ananny and Crawford 2018). They operate with millions of
data points and predict outcomes using opaque self-learning algorithms.
Most systems are so complex that even developers and researchers
sometimes fail to understand how the machine came to a specific
conclusion (Burrell 2016; Diakopoulos 2016). Accordingly, the high
complexity of such systems may lead to a lack of comprehension,
especially among the broad public with little technical knowledge and
expertise about the underlying technology (Fine Licht and Fine Licht
2020). Consequently, it may prove challenging to comprehend data-driven
decisions fully.

In such situations, trust becomes an essential factor that influences
the formation of attitudes and decision-making. A prominent definition
of trust - that is also adopted by AI researchers (e.g., Glikson and
Woolley 2020) - is the one of Mayer, Davis, and Schoorman (1995), who
define trust as ``the willingness of a party to be vulnerable to the
actions of another party based on the expectation that the other will
perform a particular action important to the trustor, irrespective of
the ability to monitor or control that other party'' (p.~712). Thus,
recognizing the complexity of the systems, people are often put in
situations where they have to rely on the decisions of ADM while not
being able to check and verify themselves whether the decision-making
process and the ultimate solution are exemplary.

Hence, a common goal of researchers and politicians is to create systems
that are trustworthy. For instance, the European approach towards AI set
by a high-level expert group actively strives for trustworthy AI design
that is also applicable to ADM systems (European Commission 2019).
According to the EU guidelines, trustworthiness can be achieved through
the fulfillment of seven ethical principles and resulting key
requirements for trustworthy intelligent systems: human oversight,
technical robustness and safety, privacy, transparency, fairness,
societal and environmental well-being, and accountability (for an
overview of global ethical guidelines, see Jobin, Ienca, and Vayena
2019). The main intention is to strengthen the public trust in such
systems that subsequently will lead to acceptance of their
implementation.\footnote{We note that we do not research the interaction
  of the fulfillment of ethical principles in order to achieve
  trustworthy AI in this paper. However, a recent paper by Kieslich,
  Keller, and Starke (2021) explores the relative importance of those
  ethical principles in the public's eye. In our paper, we solely focus
  on the effect of trust perceptions of ADM compared with HDM.}

Empirical research shows that trust is a driver for positive opinions
about technology acceptance (Hoff and Bashir 2015). Moreover, Shin
(2021a) and Shin (2021b) found evidence that trust in algorithms
positively influences perceptions of algorithmic performance.
Additionally, Shin and Park (2019) report that people who show high
trust in algorithms evaluate algorithms more positively in terms of
satisfaction and usefulness than people who show lower trust in
algorithms. Experimental research by Robinette et al. (2016) suggests
that participants followed algorithmic instructions given by a robot in
a high-risk situation due to (over)trust, even after seeing it make
mistakes. Accordingly, we hypothesize as follows:

\emph{H1. The trust in ADM will be positively related to accepting an
ADM as a viable solution for distributing COVID-19 vaccines.}

\hypertarget{social-preferences-in-the-evaluation-of-the-distribution-process}{%
\subsection{Social Preferences in the Evaluation of the Distribution
Process}\label{social-preferences-in-the-evaluation-of-the-distribution-process}}

However, the general assessment of ADM's viability is only a tiny puzzle
piece of the social challenge of vaccine distribution. A more
significant issue looms when it comes to the actual decision-making and
its results. Just because people perceive decision-making as viable in
general, a resulting decision itself is still individually evaluated and
may subsequently be questioned due to various reasons.

After all, even if an ADM system consistently arrives at formally fair
decisions, this would not automatically mean that the decisions are
generally endorsed. In this regard, ADM decisions are not different from
decisions made by humans. For instance, people may still call into
question the algorithmic systems' decisions despite their formal
attainment of optimization goals: Be it that the decisions are
unintuitive, or incomprehensible, or opposed to an individual's social
preferences and moral ideas (Brown et al. 2019; Grgic-Hlaca et al.
2018).

When it comes to the actual distribution of limited public goods (e.g.,
the distribution of vaccines against the coronavirus), decisions that
favor one social group over another may thus prove as a problem. The
literature on the assessment of distribution problems has repeatedly
shown that people exhibit not only material self-interest in their
evaluation of decisions but also social preferences. ``A person exhibits
social preferences if the person not only cares about the material
resources allocated to her but also cares about the material resources
allocated to relevant reference agents'' (Fehr and Fischbacher 2002,
C2).

Ultimately, when applying the questions by Berendt (2019) mentioned
above to the given case of allocation of vaccinations, it is likely that
different results may occur, and hence, different ADM systems can be
developed and deployed. If one aims to identify the persons highest at
risk for catching the coronavirus, particularly vulnerable groups are
prisoners and teachers (Burki 2020; Gaffney, Himmelstein, and
Woolhandler 2020; Kahn et al. 2020). Thus, an ADM system may easily
derive a solution that treats both groups equally or may even prioritize
one group over the other.

However, it has been demonstrated by prior research (Fallucchi,
Faravelli, and Quercia 2021; McKneally and Sade 2003) that such
decisions about the allocation of medical resources in the
prioritization of different social groups will be questioned and
perceived as illegitimate as people regard them as objectionable on
moral grounds. For instance, several studies found that patients'
characteristics and lifestyles influenced public perception on who to
treat first concerning organ transplantation. People would allocate
significantly less medical treatment to smokers, persons with high
alcohol consumption, or promiscuous behavior (Furnham, Ariffin, and
McClelland 2007; Huynh, Furnham, and McClelland 2020; Ubel et al. 2001).
Personal life choices can lead to the preference of one affected group
over another in the eyes of the public.

Consequently, potential decision outcomes of ADM applications need to
come into focus, especially those that may be perceived as
controversial. In this study, we do not wish to disentangle the specific
motivations for a social preference given a particular decision. Our
investigation instead focuses on the consequences of decisions that
violate social preferences. Despite the best intentions, ADM, as well as
HDM, may frequently result in controversial and unpopular decisions.

Concerning the allocation of scarce medical resources, studies showed
that, in the case of the corona pandemic, the public prioritized
treatment of younger, respectively sickest patients (Grover, McClelland,
and Furnham 2020; Huseynov, Palma, and Nayga 2020). Transferred to the
application of our study: Especially when public sentiment suggests that
unfavored groups should not be receiving any advantages, decisions
regarding vaccine distribution that are perceived as unfavorable by the
public may be seen as illegitimate. For instance, prisoners are being
punished for a crime they committed and are subsequently often
stigmatized and disadvantaged (Falk, Walkowitz, and Wirth 2009;
Kjelsberg, Skoglund, and Rustad 2007), especially in contrast to
teachers that enjoy high esteem with the majority of the German
population (dbb beamtenbund und tarifunion 2020). Consequently,
decisions that favor a group with lower social prestige compared with a
group of high social prestige may be publicly questioned as the need and
merits of the latter group are rated as higher as that of the former
group - irrespective of the algorithmic conclusions aimed at
optimization that drove the decision-making in the first place.
Therefore, it is assumed that in cases where decisions favor groups that
are of lower social prestige, the disapproval of early vaccination of
the respective group with the general public will result in lower
legitimacy of the decision. Accordingly, we hypothesize as follows:

\emph{H2. The disapproval of early vaccination of a social group will be
negatively related to the legitimacy of early vaccination of the
respective group.}

\hypertarget{the-moderating-role-of-trust-in-the-agent-making-the-decision}{%
\subsection{The Moderating Role of Trust in the Agent Making the
Decision}\label{the-moderating-role-of-trust-in-the-agent-making-the-decision}}

Concerning the importance of trust in the evaluation of ADM discussed
above, trust may not only be an explanatory variable when it comes to
general perceptions of the viability of ADM applications. Trust may more
specifically be a decisive factor in a situation in which people
encounter a decision by an agent that a) is not fully comprehensible to
them and b) is objectionable to them in its outcome. Trust may then be a
deciding factor as people that show higher trust may still perceive a
decision as legitimate even though they do not prefer the outcome.
People with lower trust in the agent making the decision will not be
affected by their lack of trust and will perceive the decision as
illegitimate.

Empirical research showed that trust in algorithms moderated the effect
of transparency, fairness, and accountability perceptions on
satisfaction with an algorithm; for people with a high trust level, the
positive effect of the relationship between the perception and
satisfaction of the ethical principles were higher than for those people
with low trust (Shin and Park 2019). Another study of Ye et al. (2019),
focusing on adopting AI in medicine in China, found that trust in AI and
medical staff negatively moderated the effect of perceived usefulness on
intention to use the respective technology. Hence, we hypothesize as
follows:

\emph{H3. Trust in the agent making the decision will moderate the
negative relation between disapproval of early vaccination of a social
group and the legitimacy of early vaccination, such that this negative
relationship will be weaker when trust in the agent making the decision
is higher.}

\hypertarget{differences-between-automated-decision-making-and-human-decision-making}{%
\subsection{Differences Between Automated Decision-making and Human
Decision-making}\label{differences-between-automated-decision-making-and-human-decision-making}}

The general impetus to implement ADM systems is to arrive at better
decisions than human decision-making (König and Wenzelburger 2021). For
instance, the use of ADM in public administration is often expected to
be superior, namely faster and cheaper, but also more reliable,
impartial, and objective than HDM (Wirtz and Müller 2018). However, even
if that were the case, the public assessment of important decisions may
deviate for various reasons, as already suggested above. Despite the
best intentions, decisions by ADM, as well as by HDM,-- while
technically correct and optimally aiming at the desired results -- may
still be negatively perceived.

In this regard, two contrasting strands of the literature are
highlighted concerning the acceptance or rejection of algorithms and
algorithmic advice, respectively, compared to human judgment:
Algorithmic aversion (Dietvorst, Simmons, and Massey 2015; Dietvorst and
Bharti 2020) and algorithmic appreciation (Logg, Minson, and Moore
2019). Notably, the research object of those studies are algorithms that
cannot be perfect in their predictions, which always come with some
degree of uncertainty. Such algorithms are used daily for recommendation
purposes or forecasting tasks.

Algorithmic aversion studies mostly argue that algorithms are unfavored
compared to humans, even if they perform better. Seminal work was done
by Dietvorst and colleagues, who found empirical evidence that people
reject algorithms when they have seen them perform and making a mistake
(Dietvorst, Simmons, and Massey 2015). This finding persisted when
participants directly compared an algorithm that factually made better
decisions than a human. In another study, Dietvorst and Bharti (2020)
argue that algorithmic aversion is correlated with the uncertainty of a
given situation; that means that a problem cannot be solved
deterministically, but can only be derived by a system, e.g.~the
prediction of stock market prices. The higher the uncertainty of a
situation, the more algorithms are rejected.

On the other side, Logg, Minson, and Moore (2019) found that laypeople
predominantly tend to follow the advice of algorithms more than the
advice of non-expert humans. However, this algorithm appreciation
vanished when a human gave advice or when they have to choose between
their one prediction and an algorithmic one. These findings are
supported by Thurman et al. (2019), who tested algorithmic appreciation
for the use case of news recommendation and found that algorithmic
recommendation was preferred to expert recommendation.

Thus, several factors seem to play a role regarding the acceptance or
rejection of algorithms, especially if a comparison is drawn to human
decision-making. Firstly, the context in which an algorithm is used is
of importance. Studies suggest that uncertainty of a situation can lead
to different degrees of algorithmic acceptance. Secondly, the role of
the human to which an algorithm is compared plays a crucial role. If
human decision-makers or advisers are considered experts, they are
mostly preferred over algorithms (even if they make worse decisions).
However, this is not true for all contexts.

In our study, we argue that if an ADM makes such a decision on vaccine
distribution, the negative effect of trust will be weaker in comparison
to human decision-making. This is because we consider vaccine
distribution to be a high-risk situation in which it has been shown that
people show overreliance on algorithmic advice (Robinette et al. 2016).
Accordingly, we hypothesize:

\emph{H4. The type of agent making the decision moderates the
interaction effect of the disapproval of vaccination of a social group
and trust in the agent making the decision on the legitimacy of the
decision, such that the negative relationship between disapproval of
vaccination and legitimacy of early vaccination is weaker for ADM making
the decision when trust in ADM is high compared to humans making the
decision when trust in humans is high.}

Figure 1 shows the conceptual model for the three hypotheses H2, H3, and
H4.

\begin{figure}
\includegraphics[width=1\linewidth]{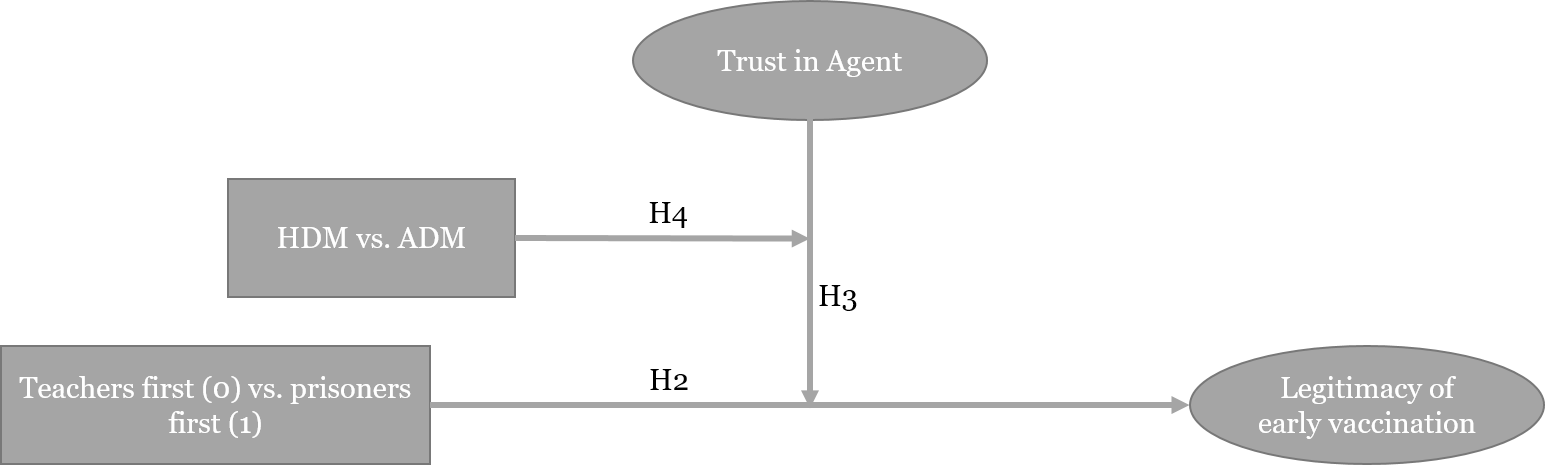} \caption{Conceptual Model for H2-H4}\label{fig:unnamed-chunk-1}
\end{figure}

\hypertarget{method}{%
\section{Method}\label{method}}

To answer the research question and hypotheses, we conducted a
cross-sectional factorial survey using a questionnaire with standardized
response options. To assess the findings, we performed the data analysis
in R (version 4.0.3) using the packages \emph{lavaan} (Rosseel 2012) and
\emph{semTools} (Jorgensen et al. 2019). We pre-registered our research
question, hypotheses as well as the measurement of the variables
(\url{https://osf.io/xhvwr}).

\hypertarget{procedure-and-survey-design}{%
\subsection{Procedure and survey
design}\label{procedure-and-survey-design}}

For screening purposes, respondents first had to indicate some
demographic information. Afterward, the respondents answered questions
concerning their opinions on the current coronavirus pandemic,
especially on the political handling of the corona situation and their
opinions on the current state and progress of vaccination. We also
included a question asking for a hypothetical vaccination prioritization
of different social groups as well as trust in the standing commission
on vaccination (STIKO), which is in charge of recommending vaccine
prioritization in Germany. Next, after assessing knowledge of artificial
intelligence (AI)\footnote{As discussed above ADM-systems may be
  regarded as a form of Artificial Intelligence (AI). Consequently, some
  questions used in the questionnaire link to the terminology of AI. On
  the one hand, it arguably is a more familiar term for the German
  public than ADM. On the other hand, we aimed for measuring some
  attitudes concerning the technology on a broader level.}, participants
were given a brief explanation of the term AI. After that, they answered
questions regarding their attitudes and opinions on AI. In the
following, participants were introduced to the use case - vaccination
distribution through an ADM system. Thereby, ADM was explained as some
form of AI. Respondents rated their trust in such a system before they
were confronted with the experimental condition.

Each participant was presented with one out of four possible scenarios,
following a 2x2 design. Participants were told that I) \emph{either} an
ADM system \emph{or} the STIKO (as human commission making decisions -
HDM) set up a vaccination distribution plan with the result that II)
\emph{either} teachers \emph{or} prisoners would be prioritized.
Following up, respondents rated the output legitimacy of the decision as
well as their fairness perception of the distribution process. To
conclude, participants were thanked, debriefed, and redirected to the
provider of the OAP where they received monetary compensation for
participation.

\hypertarget{sample}{%
\subsection{Sample}\label{sample}}

Participants were recruited with the online access panel (OAP) of the
market research institute respondi that is certified according to ISO
26362. To avoid overrepresentation and skew in the sample composition
quotas were used as a stopping rule. Survey field time was between
March, 26 and April 12, 2021. At this time, vaccination against Covid-19
in Germany was not open for anybody, but dependent on predefined risk
groups by the STIKO.

Altogether 12000 respondents from the OAP were invited to participate in
the survey. The questionnaire was accessed by 3359 persons and 3048
persons started answering the questionnaire. Of those 1184 persons were
screened out as their respective quotas were already exhausted or they
were not eligible for our survey as they did not belong to the
investigated population. At last, 1740 respondents completed the
questionnaire successfully. The dropout rate was 6.1\%, and dropouts
were equally distributed over all pages of the questionnaire.
Additionally, we filtered out those participants who answered the
questionnaire in less than 4 minutes and 30 seconds. In a pre-test, the
authors determined this as the minimum amount of time to reasonably
answer the questionnaire. The final sample consists of 1602
participants.

The average age was 48.02 (\emph{SD}=15.17). Altogether, 814 (50.8\%)
respondents identified as women and 788 (49.2\%) as men. Further, 512
(32.0\%) respondents report basic educational attainment, 536 (33.5\%)
report medium educational attainment and 554 (34.6\%) hold a degree from
higher education.

\hypertarget{measurement}{%
\subsection{Measurement}\label{measurement}}

\emph{Approval of early vaccination for a social group.} For the
measurement of preference for early vaccination of a social group,
respondents were confronted with a list of social groups, including
\emph{teachers} as well as \emph{prisoners in the closed penal system}
(full item wordings can be found in the appendix). For each group,
respondents had to rate how they would like the idea that this specific
group received prioritization for an early vaccination against the
coronavirus on a five-point Likert scale (1=do not like; 5=like,
-1=cannot judge). A Welch two Sample t-test shows that an early
vaccination of teachers (\emph{M}=4.31, \emph{SD}=1.04) was
significantly more preferred over an early vaccination of prisoners in
the closed penal system (\emph{M}=2.09, \emph{SD}=1.3),
\emph{t}(2549.8)=-49.83, \emph{p}=0.

\emph{General trust in ADM.} The general trust in ADM was measured via
four items on a five-point Likert scale ranging from 1=do not agree at
all to 5=totally agree. While the underlying construct is called
\emph{general trust in ADM} the question wordings addressed systems of
artificial intelligence. We used this approach as a) we assumed a
greater familiarity of respondents with the term of \emph{artificial
intelligence} compared with \emph{automated decision-making} and b) the
tested scales used for the assessment of our constructs were adopted
from similar research contexts that predominantly referred to AI. The
scale was adapted from the measurement of trust in recommender AI
proposed by Shin (Shin 2021a) and the used items read as follows:

\begin{itemize}
\tightlist
\item
  ``I trust that AI systems can make correct decisions.''
\item
  ``I trust the decisions made by AI systems.''
\item
  ``Decisions made by AI systems are trustworthy.''
\item
  ``I believe that decisions made by AI systems are reliable.''
\end{itemize}

The four indicators suggest good factorial validity (see table 3).

\emph{Viability of ADM for vaccine distribution.} Assessing the
perceived viability of ADM for vaccine distribution respondents had to
rate three statements on a five-point Likert scale ranging from 1=do not
agree at all to 5=totally agree.

\begin{itemize}
\tightlist
\item
  ``Computer-based decision systems are useful for the vaccine
  distribution process.''
\item
  ``I support the use of computer-based decision systems in the vaccine
  distribution process.''
\item
  ``The use of computer-based decision systems for vaccine distribution
  would help solve the problems of vaccine distribution.''
\end{itemize}

The three indicators suggest good factorial validity (see table 3).

\emph{Trust in the agent (ADM/HDM) making decisions for vaccine
distribution.} Trust in ADM for vaccine distribution was equally
measured as the general trust in ADM mentioned above except that we
changed the word ``AI'' to ``a computer system in the vaccine
distribution,'' respectively ``the STIKO in the vaccine distribution.''

Before assessing group differences using latent factor modeling, the
necessary measurement invariance of the indicators (Putnick and
Bornstein 2016) is examined by the following stepwise procedure. A first
model assessed configural invariance (M1). In a second model (M2) we
check for metric invariance by constraining the factor loadings and
comparing the two models using a \(\chi^2\)-difference-test. A
non-significant \(\chi^2\)-difference-test suggests that the model with
equality constraints does not fit worse than the model without such
constraints and the respective model parameters are considered to be
equal. Afterward, a third model (M3) with constrained indicator
intercepts is used to check for scalar invariance by also comparing it
to M2 using a \(\chi^2\)-difference-test. A model that passes this test
for measurement invariance suggest strong factorial invariance. In a
final step, constraining the residual variances of the indicators in a
fourth model (M4) we test for residual invariance.

Table 1 suggests that there is factorial invariance for the measurement
of trust in the agent making decisions for vaccine distribution. The
four indicators suggest good factorial validity (see table 3).

\begin{table}[H]

\caption{\label{tab:unnamed-chunk-3}Measurement Invariance Trust}
\centering
\resizebox{\linewidth}{!}{
\begin{tabular}[t]{llllllll}
\toprule
  & $\chi^2$ (df) & TLI & RMSEA (90\% CI) & Model comp & $\Delta$$\chi^2$ ($\Delta$df) & $\Delta$TLI & $\Delta$RMSEA\\
\midrule
M1: Configural Invariance & 10.73* (4) & 1.00 & 0.05 (0.01-0.08) &  &  &  & \\
M2: Metric Invariance & 13.76 (7) & 1.00 & 0.03 (0.00-0.06) & M1 & 3.03 (3) & 0.00 & -0.02\\
M3: Scalar Invariance & 18.94* (10) & 1.00 & 0.03 (0.01-0.06) & M2 & 5.18 (3) & 0.00 & 0.00\\
M4: Residual Invariance & 31.26* (14) & 1.00 & 0.04 (0.02-0.06) & M3 & 12.33* (4) & 0.00 & 0.01\\
\bottomrule
\multicolumn{8}{l}{\textsuperscript{a} Note. N=1602. *p<.05}\\
\end{tabular}}
\end{table}

\emph{Legitimacy of the decision for vaccination prioritization.}
Legitimacy of the decision was measured with four items on a five-point
Likert scale ranging from 1=do not agree at all to 5=totally agree. An
exemplary item was ``I accept the decision.'' The scale items were
adopted from Starke and Lünich (2020) and read as follows:

\begin{itemize}
\tightlist
\item
  ``I accept the decision.''
\item
  ``I agree with the decision.''
\item
  ``I am satisfied with the decision.''
\item
  ``I recognize the decision.''
\end{itemize}

A test for measurement invariance suggests strong factorial invariance
of the indicators measuring the legitimacy of the decision (see table
2). The four indicators suggest good factorial validity (see table 3).

Accordingly, when using the latent factors of \emph{trust in the agent
making the decision} and \emph{legitimacy of the decision for
vaccination prioritization} in the structural regression models of the
analysis, because of factorial invariance equality constraints between
the groups will be imposed on the factor loadings and the indicator
intercepts.

\begin{table}[H]

\caption{\label{tab:unnamed-chunk-4}Measurement Invariance Legitimacy}
\centering
\resizebox{\linewidth}{!}{
\begin{tabular}[t]{llllllll}
\toprule
  & $\chi^2$ (df) & TLI & RMSEA (90\% CI) & Model comp & $\Delta$$\chi^2$ ($\Delta$df) & $\Delta$TLI & $\Delta$RMSEA\\
\midrule
M1: Configural Invariance & 134.23* (8) & 0.94 & 0.20 (0.17-0.23) &  &  &  & \\
M2: Metric Invariance & 146.93* (17) & 0.97 & 0.14 (0.12-0.16) & M1 & 12.71 (9) & 0.03 & -0.06\\
M3: Scalar Invariance & 157.65* (26) & 0.98 & 0.11 (0.10-0.13) & M2 & 10.72 (9) & 0.01 & -0.03\\
M4: Residual Invariance & 214.38* (38) & 0.98 & 0.11 (0.09-0.12) & M3 & 56.73* (12) & 0.00 & 0.00\\
\bottomrule
\multicolumn{8}{l}{\textsuperscript{a} Note. N=1602. *p<.05}\\
\end{tabular}}
\end{table}

\begin{table}[H]

\caption{\label{tab:unnamed-chunk-5}Reliabiltiy Values}
\centering
\begin{tabular}[t]{>{\raggedright\arraybackslash}p{2cm}>{\raggedleft\arraybackslash}p{2cm}>{\raggedleft\arraybackslash}p{2cm}>{\raggedleft\arraybackslash}p{2cm}>{\raggedleft\arraybackslash}p{2cm}r}
\toprule
  & General Trust in ADM & ADM as Viabale Solution & Trust in Agent (ADM) & Trust in Agent (Human) & Legitimacy\\
\midrule
alpha & 0.95 & 0.94 & 0.97 & 0.97 & 0.95\\
omega & 0.95 & 0.94 & 0.97 & 0.97 & 0.95\\
omega2 & 0.95 & 0.94 & 0.97 & 0.97 & 0.95\\
omega3 & 0.95 & 0.94 & 0.97 & 0.97 & 0.95\\
avevar & 0.84 & 0.85 & 0.90 & 0.90 & 0.82\\
\bottomrule
\end{tabular}
\end{table}

\hypertarget{results}{%
\section{Results}\label{results}}

\emph{Viability of ADM for vaccine distribution.} Addressing RQ1 we ran
a latent factor analysis. In this analysis and the following, effect
coding was used for factor scaling, a procedure that ``constrains the
set of indicator intercepts to sum to zero for each construct and the
set of loadings for a given construct to average 1.0'' (Little, Slegers,
and Card 2006, 62). The eventual factor is scaled like the indicators,
which especially in the case at hand helps with interpretation. As there
were three indicators, the model is fully identified and there are no
degrees of freedom and no model fit.

Given the measurement on a five-point Likert scale, the mean of the
latent factor (\emph{M}=2.88, \emph{SD}=1.14, \emph{CI-95}(2.82; 2.94))
suggests that on average the respondents were undecided whether ADM is
to be seen as a viable solution for the distribution of the vaccine
(RQ1). All in all, there was no outright endorsement or rejection of ADM
systems for vaccine distribution.

\emph{Relationship between the general trust in ADM and the viability of
ADM for vaccine distribution.} To test the H1 of a positive relationship
between the general trust in ADM and the viability of ADM for vaccine
distribution, a structural regression model was tested that included
both constructs as latent factors. The model shows good fit
(\(\chi^2\)(13)=28.38, \emph{p}=0.01; \emph{RMSEA}=0.03
\emph{CI}{[}0.01, 0.04{]}; \emph{TLI}=1).

The parameter estimate of the regression coefficient suggests a
significant and strong effect of trust in AI on the perceived viability
of ADM for vaccine distribution (\(\beta\)=0.67, \emph{SE}= 0.03,
\emph{p}=0, \(\beta\)\textsubscript{standardized}=0.56). Accordingly, H1
is accepted.

\emph{Relationship between the disapproval of a social group's
vaccination prioritization and the legitimacy of early vaccination.} To
test H2, we estimated a structural regression model. This model includes
the factorial survey condition as an independent variable using a dummy
coded predictor (`vaccinate teachers first' = 0 vs.~`vaccinate prisoners
first' = 1). The model shows good fit (\(\chi^2\)(17)=135.38,
\emph{p}=0; \emph{RMSEA}=0.09 \emph{CI}{[}0.08, 0.11{]};
\emph{TLI}=0.98). The inadequate fit suggested by the RMSEA may be
attributed to the model's few degrees of freedom (Kenny, Kaniskan, and
McCoach 2015).

The parameter estimate of the regression coefficient suggests a
significant medium negative effect of the factorial predictor on the
perceived legitimacy of the decision (\(\beta\)=-0.66, \emph{SE}=0.06,
\emph{p}=0, \(\beta\)\textsubscript{standardized}=-0.28). That means
that the decision first to vaccinate a non-preferred group was judged as
less legitimate than the decision first to vaccinate a group where early
vaccination was generally preferred. Accordingly, H2 is accepted.

\emph{Moderation effect of trust in the agent making the decision.} H3
assumes that the trust in the agent making the decision will moderate
the relation between preference and decision legitimacy. More
specifically, we expected that this negative relation will be weaker
when trust in the agent is high.

To test H3 and subsequently H4, we again estimated a structural
regression model. This model includes as independent variables the
factorial survey condition as a dummy coded predictor (`vaccinate
teachers first' = 0 vs.~`vaccinate prisoners first' = 1) and the trust
in the agent making the decision. Additionally, a latent factor serving
as the moderator variable was estimated based on indicators calculated
as the products of the condition variable and the trust indicators using
the \texttt{indProd}-function from the package \texttt{semTools}
(Jorgensen et al. 2019).

The model shows good fit (\(\chi^2\)(60)=160.97, \emph{p}=0;
\emph{RMSEA}=0.03 \emph{CI}{[}0.03, 0.04{]}; \emph{TLI}=0.99).

The parameter estimate of the moderators regression coefficient suggests
no significant effect of the moderator variable on the perceived
legitimacy of the decision (\(\beta\)=0.02, \emph{SE}=0.05,
\emph{p}=0.74, \(\beta\)\textsubscript{standardized}=0.01). That means,
that trust in the agent making the decision had no moderating effect on
the relationship between the disapproval of early vaccination of a
social group and the legitimacy of a decision for early vaccination.
Accordingly, H3 is rejected.

\emph{Moderation effect of the agent making the decision.} H4 assumes a
difference between a condition in which \emph{either} ADM \emph{or} HDM
make decisions about early vaccination, in that the negative
relationship between disapproval of vaccination and legitimacy of early
vaccination is weaker for ADM making the decision when trust in ADM is
high compared to humans making the decision when trust in humans is
high.

We estimated a structural regression model identical to the model
estimated for H3. To assess the difference of the parameter estimates of
the moderation, however, performing multi group analysis this model
compares the two groups in which \emph{either} an ADM system decided
about vaccine prioritization \emph{or} humans (i.e., the STIKO).

The model shows good fit (\(\chi^2\)(141)=316.58, \emph{p}=0;
\emph{RMSEA}=0.04 \emph{CI}{[}0.03, 0.05{]}; \emph{TLI}=0.99).

The parameter estimate of the moderators regression coefficient shows no
moderating effect of trust in the agent making the decision in the ADM
condition (\(\beta\)=-0.03, \emph{SE}=0.07, \emph{p}=0.65,
\(\beta\)\textsubscript{standardized}=-0.02) and in the HDM condition
(\(\beta\) = 0.07, \emph{SE}=0.07, \emph{p}=0.36,
\(\beta\)\textsubscript{standardized}=0.03).

Furthermore, a test for parameter differences suggests that there is no
significant difference of the moderating effect of trust between the two
conditions (\(\beta\)=0.1, \emph{SE}=0.1, \emph{p}=0.34,
\(\beta\)\textsubscript{standardized}=0.05). H4 is also rejected.

\hypertarget{discussion}{%
\section{Discussion}\label{discussion}}

In focusing on AI implementation against one of the biggest current
challenges for humanity, namely Covid-19, our study adds to the current
research of a hotly debated social issue. As AI applications are already
in extensive use that will most likely increase over the coming years,
it is crucial to understand how the public perceives their widespread
deployment, especially in high-risk situations. Here, we mainly focused
on the role of trust and its effect on the legitimacy of publicly
preferred vs.~unpreferred solutions.

The results of the factorial survey suggest that the German public is
altogether indifferent about ADM usage to allocate vaccination against
the coronavirus. Answering our research question, the individual
technological approach to tackle this important current issue is not
rejected but also not overly welcomed by German citizens. This insight
is in line with research that suggests that while German citizens are
generally in favor of AI (bitkom 2018), they often show little interest
in AI and specific use cases (Meinungsmonitor Künstliche Intelligenz
2021). Overall, there is low involvement of the German public regarding
the actual implementation of ADM systems.

In confirming H1, we see that trust in ADM leads to greater acceptance
of the use of ADM in the allocation of coronavirus vaccines. This
finding is also consistent with previous research showing that trust
positively affects perceived satisfaction and usefulness of ADM systems.
Hence, building trust in ADM systems proves to be a fruitful way to
legitimatize AI use in public administration decision-making.
Consequently, it may be assumed that efforts to promote the use of ADM
systems in the management of current crises fall on open ears,
especially with people who are generally in favor of the respective
innovations and who show considerable trust in their beneficial
potential.

However, as initially well-received deployments may lead to unpopular
and consequence-laden outcomes, we subsequently contrasted vaccine
allocation decisions of high public preference with decisions of low
public preference. Our findings reveal that ethical considerations might
not be in line -- or even strongly oppose -- public preferences. For
instance, prisoners are at high risk of the coronavirus (Burki 2020).
However, public sentiment strongly opposes the idea of prioritizing the
respective group. This disapproval of early vaccination for an unpopular
social group is negatively related to the legitimacy of early
vaccination for the respective group.

These findings correspond to the literature on the allocation of scarce
medical resources. Personal characteristics and life choices affect
social preferences and influence how the public legitimates a
prioritization of respective groups. Prisoners are being punished for a
crime they committed, and the social preference of such persons is low
in the German population, especially in contrast to teachers. Hence,
public preference depends on the specific social characteristics the
respective groups possess (Luyten, Tubeuf, and Kessels 2020; Sprengholz
et al. 2021). Existing studies on the allocation of scarce resources
concerning Covid-19 often do not differentiate between the groups
affected but rather on the ethical ground principles on which decisions
are based (Huseynov, Palma, and Nayga 2020; Grover, McClelland, and
Furnham 2020). Thus, further studies should elaborate on our findings
and probe into different preference patterns among the public to
mitigate the detrimental effects of unpopular decisions on accepting ADM
systems.

In a subsequent step, we asked whether trust moderates the link between
social preferences and legitimacy. After all, trusting someone to make
the right call may help to accept an otherwise unpopular decision.
Contrary to expectations, in situations of significant discrepancy
between expectations and actual outcomes, trust does not moderate the
effect of social group preference on legitimacy. Furthermore, there was
no difference between ADM and HDM. This finding has far-reaching
implications. Based on the respective goal formulation, algorithms are
expected to produce accurate and objective results. On the one hand,
respective ADM systems are supposed to arrive at ethically sound
decisions (e.g., as required by the high-level expert group of the
European Commission 2019). On the other hand, correct and ethically
tenable outcomes may not be in line with the opinions of the broad
public. As the overarching goal is to build trustworthy AI systems, this
points to a potential major conflict as not all these demands may be met
satisfactorily. Hence, we show that trustworthy AI may not be the
solution to every ethical problem in the eye of the public. As ADM gets
integrated in more and more parts of societal life, it is crucial to
have these findings in mind. We are far away from a point, where people
trust wholeheartedly rely on the decisions of a machine. Legitimacy is
first and foremost influenced, at least in our case, by public
preferences of the solution an agent proposes.

\hypertarget{implications}{%
\section{Implications}\label{implications}}

While we highly welcome the necessity of ethical AI guidelines, we
observe that ADM decisions and demands for trustworthy AI may sometimes
not be in line but direct conflict with public perceptions of AI's
output. Thus, alongside the development of ethical AI in technical
terms, companies and researchers also have to acknowledge the relevance
of public opinion. As seen in the case of vaccine distribution in the US
(Guo and Hao 2020) and Germany that often created false, unexpected, and
unpopular results, particular outcomes may backfire and fuel public
outrage against the use of ADM. Hence, decision-makers must weigh
ethical considerations and the public's will in light of probable public
resistance against ADM decisions.

As another potential remedy to the detected dilemma, studies focusing on
\emph{Explainable AI} (XAI) highlight the importance of explaining ADM
decisions to citizens (for an overview, see Miller 2019). Empirical
studies found that explaining ADM decisions leads to greater trust in
those systems and, in turn to greater acceptance (Shin 2021a). Thus,
further studies could enhance our design and test if the more or less
detailed and comprehensible explanation for a decisive outcome would
soften the negative effect of social group preference on decision
legitimacy. After all, the conflict between ethical decisions and their
negative public perception in light of public opinion may be mitigated
with specific communicative strategies involving convincing explanations
that make the inner workings of ADM comprehensible to a lay audience.

\hypertarget{conclusion}{%
\section{Conclusion}\label{conclusion}}

The vaccination program against the novel coronavirus currently poses a
challenge of global dimension and, as such, is the subject of a
controversial social debate. Decision-makers have to allocate scarce
medical resources considering many factors, including practical and
moral questions but also in consideration of public opinion. ADM systems
are deployed to support this process in providing suggestions or even
autonomously deciding upon the rank order for vaccination.

Our research suggests that, generally, the usage of ADM in combating the
coronavirus pandemic is perceived ambivalently as a viable strategy with
the German public and that the general trust in AI is an essential
driver of such viability perceptions. However, irrespective of actual
discrimination -- be it necessary or faulty -- by ADM, we show that as
soon as publicly unpreferred decisions regarding the allocation of
vaccines are proposed, these decisions are perceived as less legitimate.
We subsequently inquired about the moderating role of trust in agents
making decisions on the legitimacy of unpreferred decisions in the
allocation process. Contrary to expectations, the trust in the agent
making the decision did not have the expected mitigating effect. As
there was also no difference between human decision-makers or ADM, this
raises important questions concerning the expected future deployments of
ADM in administrative decision-making.

As there are potentially many ethically correct and preferable yet
widely unpopular decisions that ADM systems will propose in the future,
we conclude that there are severe challenges for current initiatives
promoting the implementation of trustworthy AI.

\hypertarget{acknowledgement}{%
\section*{Acknowledgement}\label{acknowledgement}}
\addcontentsline{toc}{section}{Acknowledgement}

The authors thank Kira Klinger for her thoughtful feedback on the paper.

\hypertarget{funding}{%
\section*{Funding}\label{funding}}
\addcontentsline{toc}{section}{Funding}

This study was conducted as part of the project \emph{Fair Artificial
Intelligence Reasoning} (FAIR). The project is funded by the
Volkswagenstiftung.

\hypertarget{references}{%
\section*{References}\label{references}}
\addcontentsline{toc}{section}{References}

\hypertarget{refs}{}
\begin{CSLReferences}{1}{0}
\leavevmode\hypertarget{ref-Ananny.2018}{}%
Ananny, Mike, and Kate Crawford. 2018. {``Seeing Without Knowing:
Limitations of the Transparency Ideal and Its Application to Algorithmic
Accountability.''} \emph{New Media {\&} Society} 20 (3): 973--89.
\url{https://doi.org/10.1177/1461444816676645}.

\leavevmode\hypertarget{ref-Berendt.2019}{}%
Berendt, Bettina. 2019. {``AI for the Common Good?! Pitfalls,
Challenges, and Ethics Pen-Testing.''} \emph{Paladyn, Journal of
Behavioral Robotics} 10 (1): 44--65.
\url{https://doi.org/10.1515/pjbr-2019-0004}.

\leavevmode\hypertarget{ref-bitkom.2018}{}%
bitkom. 2018. {``K{ü}nstliche Intelligenz: Bundesb{ü}rger Sehen Vor
Allem Chancen.''} Edited by bitkom.
\url{https://www.bitkom.org/Presse/Presseinformation/Kuenstliche-Intelligenz-Bundesbuerger-sehen-vor-allem-Chancen}.

\leavevmode\hypertarget{ref-Bragazzi.2020}{}%
Bragazzi, Nicola Luigi, Haijiang Dai, Giovanni Damiani, Masoud
Behzadifar, Mariano Martini, and Jianhong Wu. 2020. {``How Big Data and
Artificial Intelligence Can Help Better Manage the COVID-19 Pandemic.''}
\emph{International Journal of Environmental Research and Public Health}
17 (9): 3176. \url{https://doi.org/10.3390/ijerph17093176}.

\leavevmode\hypertarget{ref-Brown.2019}{}%
Brown, Anna, Alexandra Chouldechova, Emily Putnam-Hornstein, Andrew
Tobin, and Rhema Vaithianathan. 2019. {``Toward Algorithmic
Accountability in Public Services.''} In \emph{Proceedings of the 2019
CHI Conference on Human Factors in Computing Systems}, edited by Stephen
Brewster, Geraldine Fitzpatrick, Anna Cox, and Vassilis Kostakos, 1--12.
New York, NY, USA: ACM. \url{https://doi.org/10.1145/3290605.3300271}.

\leavevmode\hypertarget{ref-Burki.2020}{}%
Burki, Talha. 2020. {``Prisons Are {`in No Way Equipped'} to Deal with
COVID-19.''} \emph{The Lancet} 395 (10234): 1411--12.
\url{https://doi.org/10.1016/S0140-6736(20)30984-3}.

\leavevmode\hypertarget{ref-Burrell.2016}{}%
Burrell, Jenna. 2016. {``How the Machine {`Thinks'}: Understanding
Opacity in Machine Learning Algorithms.''} \emph{Big Data {\&} Society}
3 (1): 205395171562251. \url{https://doi.org/10.1177/2053951715622512}.

\leavevmode\hypertarget{ref-Calandra.2020}{}%
Calandra, Davide, and Matteo Favareto. 2020. {``Artificial Intelligence
to Fight COVID-19 Outbreak Impact: An Overview: 84-104 Pages / European
Journal of Social Impact and Circular Economy, Vol 1 No 3 (2020): CSR
and Circular Economy as a Remedy for Companies Fighting Systemic Crises
/ European Journal of Social Impact and Circular Economy, Vol 1 No 3
(2020): CSR and Circular Economy as a Remedy for Companies Fighting
Systemic Crises.''} \url{https://doi.org/10.13135/2704-9906/5067}.

\leavevmode\hypertarget{ref-Cave.2019}{}%
Cave, Stephen, Kate Coughlan, and Kanta Dihal. 2019. {``{{}}Scary
Robots{{}}: Examining Public Responses to AI.''} In \emph{Proceedings of
the 2019 AAAI/ACM Conference on AI, Ethics, and Society}, edited by
Vincent Conitzer, Gillian Hadfield, and Shannon Vallor, 331--37. New
York, NY, USA: ACM. \url{https://doi.org/10.1145/3306618.3314232}.

\leavevmode\hypertarget{ref-Ciesielski.2021}{}%
Ciesielski, Rebecca, Maximilian Zierer, and Ann-Kathrin Wetter. 2021.
{``Impftermin-Vergabe: Werden {Ä}ltere Benachteiligt?''}
\url{https://www.br.de/nachrichten/bayern/impftermin-vergabe-werden-aeltere-benachteiligt,SSbbNJE}.

\leavevmode\hypertarget{ref-Crawford.2016}{}%
Crawford, Kate, Meredith Whittaker, Madeleine Elish Elish, Solon
Barocas, Aaron Plasek, and Kadija Ferryman. 2016. {``The AI Now Report:
The Social and Economic Implications of Artificial Intelligence
Technologies in the Near-Term.''}
\url{https://ainowinstitute.org/AI_Now_2016_Report.pdf}.

\leavevmode\hypertarget{ref-Dawes.1989}{}%
Dawes, R. M., D. Faust, and P. E. Meehl. 1989. {``Clinical Versus
Actuarial Judgment.''} \emph{Science} 243 (4899): 1668--74.
\url{https://doi.org/10.1126/science.2648573}.

\leavevmode\hypertarget{ref-dbbbeamtenbundundtarifunion.2020}{}%
dbb beamtenbund und tarifunion. 2020. {``Dbb b{ü}rgerbefragung
{Ö}ffentlicher Dienst: Einsch{ä}tzungen, Erfahrungen Und Erwartungen Der
b{ü}rger.''} Edited by forsa. {dbb beamtenbund und tarifunion}.
\url{https://digital.zlb.de/viewer/api/v1/records/34069248_2020/files/images/forsa_2020.pdf/full.pdf}.

\leavevmode\hypertarget{ref-Diakopoulos.2016}{}%
Diakopoulos, Nicholas. 2016. {``Accountability in Algorithmic Decision
Making.''} \emph{Communications of the ACM} 59 (2): 56--62.
\url{https://doi.org/10.1145/2844110}.

\leavevmode\hypertarget{ref-Dietvorst.2020}{}%
Dietvorst, Berkeley J., and Soaham Bharti. 2020. {``People Reject
Algorithms in Uncertain Decision Domains Because They Have Diminishing
Sensitivity to Forecasting Error.''} \emph{Psychological Science} 31
(10): 1302--14. \url{https://doi.org/10.1177/0956797620948841}.

\leavevmode\hypertarget{ref-Dietvorst.2015}{}%
Dietvorst, Berkeley J., Joseph P. Simmons, and Cade Massey. 2015.
{``Algorithm Aversion: People Erroneously Avoid Algorithms After Seeing
Them Err.''} \emph{Journal of Experimental Psychology. General} 144 (1):
114--26. \url{https://doi.org/10.1037/xge0000033}.

\leavevmode\hypertarget{ref-EuropeanCommission.2019}{}%
European Commission. 2019. {``Ethics Guidelines for Trustworthy AI.''}
\url{https://digital-strategy.ec.europa.eu/en/library/ethics-guidelines-trustworthy-ai}.

\leavevmode\hypertarget{ref-Falk.2009}{}%
Falk, Armin, Gari Walkowitz, and Wolfgang Wirth. 2009.
{``Benachteiligung Wegen Mangelnden Vertrauens? Eine Experimentelle
Studie Zur Arbeitsmarktintegration von Strafgefangenen.''}
\emph{Monatsschrift f{ü}r Kriminologie Und Strafrechtsreform} 92 (6):
526--46. \url{https://doi.org/10.1515/mks-2009-920602}.

\leavevmode\hypertarget{ref-Fallucchi.2021}{}%
Fallucchi, Francesco, Marco Faravelli, and Simone Quercia. 2021. {``Fair
Allocation of Scarce Medical Resources in the Time of COVID-19: What Do
People Think?''} \emph{Journal of Medical Ethics} 47 (1): 3--6.
\url{https://doi.org/10.1136/medethics-2020-106524}.

\leavevmode\hypertarget{ref-Fehr.2002}{}%
Fehr, Ernst, and Urs Fischbacher. 2002. {``Why Social Preferences Matter
-- the Impact of Non--Selfish Motives on Competition, Cooperation and
Incentives.''} \emph{The Economic Journal} 112 (478): C1--33.
\url{https://doi.org/10.1111/1468-0297.00027}.

\leavevmode\hypertarget{ref-FineLicht.2020}{}%
Fine Licht, Karl de, and Jenny de Fine Licht. 2020. {``Artificial
Intelligence, Transparency, and Public Decision-Making.''} \emph{AI {\&}
SOCIETY} 35 (4): 917--26.
\url{https://doi.org/10.1007/s00146-020-00960-w}.

\leavevmode\hypertarget{ref-Furnham.2007}{}%
Furnham, Adrian, Alicia Ariffin, and Alastair McClelland. 2007.
{``Factors Affecting Allocation of Scarce Medical Resources Across
Life-Threatening Medical Conditions.''} \emph{Journal of Applied Social
Psychology} 37 (12): 2903--21.
\url{https://doi.org/10.1111/j.1559-1816.2007.00287.x}.

\leavevmode\hypertarget{ref-Gaffney.2020}{}%
Gaffney, Adam W., David Himmelstein, and Steffie Woolhandler. 2020.
{``Risk for Severe COVID-19 Illness Among Teachers and Adults Living
with School-Aged Children.''} \emph{Annals of Internal Medicine} 173
(9): 765--67. \url{https://doi.org/10.7326/M20-5413}.

\leavevmode\hypertarget{ref-Glikson.2020}{}%
Glikson, Ella, and Anita Williams Woolley. 2020. {``Human Trust in
Artificial Intelligence: Review of Empirical Research.''} \emph{Academy
of Management Annals} 14 (2): 627--60.
\url{https://doi.org/10.5465/annals.2018.0057}.

\leavevmode\hypertarget{ref-GrgicHlaca.2018}{}%
Grgic-Hlaca, Nina, Elissa M. Redmiles, Krishna P. Gummadi, and Adrian
Weller. 2018. {``Human Perceptions of Fairness in Algorithmic Decision
Making.''} In \emph{Proceedings of the 2018 World Wide Web Conference on
World Wide Web - WWW '18}, edited by Pierre-Antoine Champin, Fabien
Gandon, Mounia Lalmas, and Panagiotis G. Ipeirotis, 903--12. New York,
New York, USA: {ACM Press}.
\url{https://doi.org/10.1145/3178876.3186138}.

\leavevmode\hypertarget{ref-Grover.2020}{}%
Grover, Simmy, Alastair McClelland, and Adrian Furnham. 2020.
{``Preferences for Scarce Medical Resource Allocation: Differences
Between Experts and the General Public and Implications for the COVID-19
Pandemic.''} \emph{British Journal of Health Psychology} 25 (4):
889--901. \url{https://doi.org/10.1111/bjhp.12439}.

\leavevmode\hypertarget{ref-Guo.2020}{}%
Guo, Eileen, and Karen Hao. 2020. {``This Is the Stanford Vaccine
Algorithm That Left Out Frontline Doctors.''} Edited by MIT Technology
Review.
\url{https://www.technologyreview.com/2020/12/21/1015303/stanford-vaccine-algorithm/}.

\leavevmode\hypertarget{ref-Hartmann.2021}{}%
Hartmann, Kathrin, and Georg Wenzelburger. 2021. {``Uncertainty, Risk
and the Use of Algorithms in Policy Decisions: A Case Study on Criminal
Justice in the USA.''} \emph{Policy Sciences} 54 (2): 269--87.
\url{https://doi.org/10.1007/s11077-020-09414-y}.

\leavevmode\hypertarget{ref-Heinrichs.2021}{}%
Heinrichs, Bert. 2021. {``Discrimination in the Age of Artificial
Intelligence.''} \emph{AI {\&} SOCIETY}, 1--12.
\url{https://doi.org/10.1007/s00146-021-01192-2}.

\leavevmode\hypertarget{ref-Hoff.2015}{}%
Hoff, Kevin Anthony, and Masooda Bashir. 2015. {``Trust in Automation:
Integrating Empirical Evidence on Factors That Influence Trust.''}
\emph{Human Factors} 57 (3): 407--34.
\url{https://doi.org/10.1177/0018720814547570}.

\leavevmode\hypertarget{ref-Huseynov.2020}{}%
Huseynov, Samir, Marco A. Palma, and Rodolfo M. Nayga. 2020. {``General
Public Preferences for Allocating Scarce Medical Resources During
COVID-19.''} \emph{Frontiers in Public Health} 8: 587423.
\url{https://doi.org/10.3389/fpubh.2020.587423}.

\leavevmode\hypertarget{ref-Huynh.2020}{}%
Huynh, Agathe Nguyen, Adrian Furnham, and Alastair McClelland. 2020.
{``A Cross-Cultural Investigation of the Lifestyle Factors Affecting
Laypeople's Allocation of a Scarce Medical Resource.''} \emph{Health} 12
(02): 141--57. \url{https://doi.org/10.4236/health.2020.122013}.

\leavevmode\hypertarget{ref-Jacob.2020}{}%
Jacob, Steve, and Justin Lawarée. 2020. {``The Adoption of Contact
Tracing Applications of COVID-19 by European Governments.''}
\emph{Policy Design and Practice}, 1--15.
\url{https://doi.org/10.1080/25741292.2020.1850404}.

\leavevmode\hypertarget{ref-Jobin.2019}{}%
Jobin, Anna, Marcello Ienca, and Effy Vayena. 2019. {``The Global
Landscape of AI Ethics Guidelines.''} \emph{Nature Machine Intelligence}
1 (9): 389--99. \url{https://doi.org/10.1038/s42256-019-0088-2}.

\leavevmode\hypertarget{ref-Jorgensen.2019}{}%
Jorgensen, Terrence D., Sunthud Pornprasertmanit, Alexander M.
Schoemann, and Yves Rosseel. 2019. {``semTools: Useful Tools for
Structural Equation Modeling. R Package Version 0.5-2.''}
\url{https://CRAN.R-project.org/package=semTools}.

\leavevmode\hypertarget{ref-Kahn.2020}{}%
Kahn, Benjamin, Lisa Brown, William Foege, and Helene Gayle, eds. 2020.
\emph{Framework for Equitable Allocation of COVID-19 Vaccine}.
Washington (DC). \url{https://doi.org/10.17226/25917}.

\leavevmode\hypertarget{ref-Kaufmann.2016}{}%
Kaufmann, Esther, and Werner W. Wittmann. 2016. {``The Success of Linear
Bootstrapping Models: Decision Domain-, Expertise-, and
Criterion-Specific Meta-Analysis.''} \emph{PloS One} 11 (6): e0157914.
\url{https://doi.org/10.1371/journal.pone.0157914}.

\leavevmode\hypertarget{ref-Kelley.2019}{}%
Kelley, Patrick Gage, Yongwei Yang, Courtney Heldreth, Christopher
Moessner, Aaron Sedley, Andreas Kramm, David Newman, and Allison
Woodruff. 2019. {``Happy and Assured That Life Will Be Easy 10 Years
from Now: Perceptions of Artificial Intelligence in 8 Countries.''}
\url{http://arxiv.org/pdf/2001.00081v1}.

\leavevmode\hypertarget{ref-Kenny.2015}{}%
Kenny, David A., Burcu Kaniskan, and D. Betsy McCoach. 2015. {``The
Performance of RMSEA in Models with Small Degrees of Freedom.''}
\emph{Sociological Methods {\&} Research} 44 (3): 486--507.
\url{https://doi.org/10.1177/0049124114543236}.

\leavevmode\hypertarget{ref-Kieslich.2021}{}%
Kieslich, Kimon, Birte Keller, and Christopher Starke. 2021.
{``AI-Ethics by Design. Evaluating Public Perception on the Importance
of Ethical Design Principles of AI.''}
\url{http://arxiv.org/pdf/2106.00326v1}.

\leavevmode\hypertarget{ref-Kieslich.2021b}{}%
Kieslich, Kimon, Marco Lünich, and Frank Marcinkowski. 2021. {``The
Threats of Artificial Intelligence Scale (TAI).''} \emph{International
Journal of Social Robotics}.
\url{https://doi.org/10.1007/s12369-020-00734-w}.

\leavevmode\hypertarget{ref-Kjelsberg.2007}{}%
Kjelsberg, Ellen, Tom Hilding Skoglund, and Aase-Bente Rustad. 2007.
{``Attitudes Towards Prisoners, as Reported by Prison Inmates, Prison
Employees and College Students.''} \emph{BMC Public Health} 7: 71.
\url{https://doi.org/10.1186/1471-2458-7-71}.

\leavevmode\hypertarget{ref-Konig.2021}{}%
König, Pascal D., and Georg Wenzelburger. 2021. {``Between
Technochauvinism and Human-Centrism: Can Algorithms Improve
Decision-Making in Democratic Politics?''} \emph{European Political
Science}, 1--18. \url{https://doi.org/10.1057/s41304-020-00298-3}.

\leavevmode\hypertarget{ref-Kuncel.2013}{}%
Kuncel, Nathan R., David M. Klieger, Brian S. Connelly, and Deniz S.
Ones. 2013. {``Mechanical Versus Clinical Data Combination in Selection
and Admissions Decisions: A Meta-Analysis.''} \emph{The Journal of
Applied Psychology} 98 (6): 1060--72.
\url{https://doi.org/10.1037/a0034156}.

\leavevmode\hypertarget{ref-Liang.2017}{}%
Liang, Yuhua, and Seungcheol Austin Lee. 2017. {``Fear of Autonomous
Robots and Artificial Intelligence: Evidence from National
Representative Data with Probability Sampling.''} \emph{International
Journal of Social Robotics} 9 (3): 379--84.
\url{https://doi.org/10.1007/s12369-017-0401-3}.

\leavevmode\hypertarget{ref-Little.2006}{}%
Little, Todd D., David W. Slegers, and Noel A. Card. 2006. {``A
Non-Arbitrary Method of Identifying and Scaling Latent Variables in SEM
and MACS Models.''} \emph{Structural Equation Modeling: A
Multidisciplinary Journal} 13 (1): 59--72.
\url{https://doi.org/10.1207/s15328007sem13013}.

\leavevmode\hypertarget{ref-Logg.2019}{}%
Logg, Jennifer M., Julia A. Minson, and Don A. Moore. 2019. {``Algorithm
Appreciation: People Prefer Algorithmic to Human Judgment.''}
\emph{Organizational Behavior and Human Decision Processes} 151 (10):
90--103. \url{https://doi.org/10.1016/j.obhdp.2018.12.005}.

\leavevmode\hypertarget{ref-Luyten.2020}{}%
Luyten, Jeroen, Sandy Tubeuf, and Roselinde Kessels. 2020. {``Who Should
Get It First? Public Preferences for Distributing a COVID-19 Vaccine.''}
\emph{Covid Economics, Vetted and Real-Time Papers}, no. 57: 1--19.
\url{https://dial.uclouvain.be/pr/boreal/object/boreal:238015}.

\leavevmode\hypertarget{ref-Malik.2020}{}%
Malik, Yashpal Singh, Shubhankar Sircar, Sudipta Bhat, Mohd Ikram
Ansari, Tripti Pande, Prashant Kumar, Basavaraj Mathapati, et al. 2020.
{``How Artificial Intelligence May Help the Covid--19 Pandemic: Pitfalls
and Lessons for the Future.''} \emph{Reviews in Medical Virology}.
\url{https://doi.org/10.1002/rmv.2205}.

\leavevmode\hypertarget{ref-Matrajt.2020}{}%
Matrajt, Laura, Julie Eaton, Tiffany Leung, and Elizabeth R. Brown.
2020. {``Vaccine Optimization for COVID-19, Who to Vaccinate First?''}
\emph{medRxiv}, 2020.08.14.20175257.
\url{https://doi.org/10.1101/2020.08.14.20175257}.

\leavevmode\hypertarget{ref-Mayer.1995}{}%
Mayer, Roger C., James H. Davis, and F. David Schoorman. 1995. {``An
Integrative Model of Organizational Trust.''} \emph{Academy of
Management Review} 20 (3): 709--34.
\url{https://doi.org/10.5465/amr.1995.9508080335}.

\leavevmode\hypertarget{ref-McKneally.2003}{}%
McKneally, Martin F., and Robert M. Sade. 2003. {``The Prisoner Dilemma:
Should Convicted Felons Have the Same Access to Heart Transplantation as
Ordinary Citizens? Opposing Views.''} \emph{The Journal of Thoracic and
Cardiovascular Surgery} 125 (3): 451--53.
\url{https://doi.org/10.1067/mtc.2003.61}.

\leavevmode\hypertarget{ref-MeinungsmonitorKunstlicheIntelligenz.2021}{}%
Meinungsmonitor Künstliche Intelligenz. 2021. {``What Does the Public
Think about Artificial Intelligence? How Does the Media Report on It?''}
\url{https://www.cais.nrw/en/memoki_en/}.

\leavevmode\hypertarget{ref-Miller.2019}{}%
Miller, Tim. 2019. {``Explanation in Artificial Intelligence: Insights
from the Social Sciences.''} \emph{Artificial Intelligence} 267: 1--38.
\url{https://doi.org/10.1016/j.artint.2018.07.007}.

\leavevmode\hypertarget{ref-Nguyen.2020}{}%
Nguyen, Dinh, Ming Ding, Pubudu N. Pathirana, and Aruna Seneviratne.
2020. {``Blockchain and AI-Based Solutions to Combat Coronavirus
(COVID-19)-Like Epidemics: A Survey.''} TechRxiv.
\url{https://doi.org/10.36227/techrxiv.12121962.v1}.

\leavevmode\hypertarget{ref-Putnick.2016}{}%
Putnick, Diane L., and Marc H. Bornstein. 2016. {``Measurement
Invariance Conventions and Reporting: The State of the Art and Future
Directions for Psychological Research.''} \emph{Developmental Review}
41: 71--90. \url{https://doi.org/10.1016/j.dr.2016.06.004}.

\leavevmode\hypertarget{ref-Ratcliffe.2000}{}%
Ratcliffe, Julie. 2000. {``Public Preferences for the Allocation of
Donor Liver Grafts for Transplantation.''} \emph{Health Economics} 9
(2): 137--48.
\url{https://doi.org/10.1002/(SICI)1099-1050(200003)9:2\%3C137::AID-HEC489\%3E3.0.CO;2-1}.

\leavevmode\hypertarget{ref-Robinette.2016}{}%
Robinette, Paul, Wenchen Li, Robert Allen, Ayanna M. Howard, and Alan R.
Wagner. 2016. {``Overtrust of Robots in Emergency Evacuation
Scenarios.''} In \emph{2016 11th ACM/IEEE International Conference on
Human-Robot Interaction (HRI)}, 101--8. IEEE.
\url{https://doi.org/10.1109/HRI.2016.7451740}.

\leavevmode\hypertarget{ref-Rosseel.2012}{}%
Rosseel, Yves. 2012. {``Lavaan : An r Package for Structural Equation
Modeling.''} \emph{Journal of Statistical Software} 48 (2).
\url{https://doi.org/10.18637/jss.v048.i02}.

\leavevmode\hypertarget{ref-Shin.2021}{}%
Shin, Donghee. 2021a. {``The Effects of Explainability and Causability
on Perception, Trust, and Acceptance: Implications for Explainable
AI.''} \emph{International Journal of Human-Computer Studies} 146:
102551. \url{https://doi.org/10.1016/j.ijhcs.2020.102551}.

\leavevmode\hypertarget{ref-Shin.2021b}{}%
---------. 2021b. {``Why Does Explainability Matter in News Analytic
Systems? Proposing Explainable Analytic Journalism.''} \emph{Journalism
Studies} 22 (8): 1047--65.
\url{https://doi.org/10.1080/1461670X.2021.1916984}.

\leavevmode\hypertarget{ref-Shin.2019}{}%
Shin, Donghee, and Yong Jin Park. 2019. {``Role of Fairness,
Accountability, and Transparency in Algorithmic Affordance.''}
\emph{Computers in Human Behavior} 98: 277--84.
\url{https://doi.org/10.1016/j.chb.2019.04.019}.

\leavevmode\hypertarget{ref-Sipior.2020}{}%
Sipior, Janice C. 2020. {``Considerations for Development and Use of AI
in Response to COVID-19.''} \emph{International Journal of Information
Management} 55: 102170.
\url{https://doi.org/10.1016/j.ijinfomgt.2020.102170}.

\leavevmode\hypertarget{ref-Sprengholz.2021}{}%
Sprengholz, Philipp, Lars Korn, Sarah Eitze, and Cornelia Betsch. 2021.
{``Allocation of COVID-19 Vaccination: When Public Prioritisation
Preferences Differ from Official Regulations.''} {Open Science
Framework}. \url{https://doi.org/10.17605/OSF.IO/CKHBA}.

\leavevmode\hypertarget{ref-Starke.2020}{}%
Starke, Christopher, and Marco Lünich. 2020. {``Artificial Intelligence
for Political Decision-Making in the European Union: Effects on
Citizens' Perceptions of Input, Throughput, and Output Legitimacy.''}
\emph{Data {\&} Policy} 2. \url{https://doi.org/10.1017/dap.2020.19}.

\leavevmode\hypertarget{ref-Thurman.2019}{}%
Thurman, Neil, Judith Moeller, Natali Helberger, and Damian Trilling.
2019. {``My Friends, Editors, Algorithms, and i.''} \emph{Digital
Journalism} 7 (4): 447--69.
\url{https://doi.org/10.1080/21670811.2018.1493936}.

\leavevmode\hypertarget{ref-Ubel.2001}{}%
Ubel, P. A., C. Jepson, J. Baron, T. Mohr, S. McMorrow, and D. A. Asch.
2001. {``Allocation of Transplantable Organs: Do People Want to Punish
Patients for Causing Their Illness?''} \emph{Liver Transplantation :
Official Publication of the American Association for the Study of Liver
Diseases and the International Liver Transplantation Society} 7 (7):
600--607. \url{https://doi.org/10.1053/jlts.2001.25361}.

\leavevmode\hypertarget{ref-Wirtz.2018}{}%
Wirtz, Bernd W., and Wilhelm M. Müller. 2018. {``An Integrated
Artificial Intelligence Framework for Public Management.''} \emph{Public
Management Review} 32 (5): 1--25.
\url{https://doi.org/10.1080/14719037.2018.1549268}.

\leavevmode\hypertarget{ref-Wu.2020}{}%
Wu, Katherine J., and Mike Isaac. 2020. {``Frontline Workers Were Left
Off the Vaccine List at Stanford Medical Center in Palo Alto. They
Fought Back.''} \emph{New York Times}.
\url{https://www.nytimes.com/2020/12/18/world/covid-stanford-health-center-vaccine-protest.html}.

\leavevmode\hypertarget{ref-Ye.2019}{}%
Ye, Tiantian, Jiaolong Xue, Mingguang He, Jing Gu, Haotian Lin, Bin Xu,
and Yu Cheng. 2019. {``Psychosocial Factors Affecting Artificial
Intelligence Adoption in Health Care in China: Cross-Sectional Study.''}
\emph{Journal of Medical Internet Research} 21 (10): e14316.
\url{https://doi.org/10.2196/14316}.

\leavevmode\hypertarget{ref-Zhang.2019}{}%
Zhang, Baobao, and Allan Dafoe. 2019. {``Artificial Intelligence:
American Attitudes and Trends.''} \emph{SSRN Electronic Journal}.
\url{https://doi.org/10.2139/ssrn.3312874}.

\end{CSLReferences}

\bibliographystyle{unsrt}
\bibliography{references.bib}

\end{document}